\documentclass[twocolumn,showpacs,preprintnumbers,amsmath,amssymb,floatfix]{revtex4-1}

\usepackage{graphicx}
\usepackage{dcolumn}
\usepackage{bm}
\usepackage{upgreek}
\usepackage{hyperref}

\begin{document}

\title{Zeeman Slowers for Strontium based on Permanent Magnets}

\author{Ian~R.~Hill}
\email{ian.hill@npl.co.uk}
\affiliation{National Physical Laboratory, Hampton Road, Teddington, TW11~0LW, UK}

\author{Yuri~B.~Ovchinnikov}
\affiliation{National Physical Laboratory, Hampton Road, Teddington, TW11~0LW, UK}

\author{Elizabeth~M.~Bridge}
\affiliation{National Physical Laboratory, Hampton Road, Teddington, TW11~0LW, UK}


\author{E.~Anne~Curtis}
\affiliation{National Physical Laboratory, Hampton Road, Teddington, TW11~0LW, UK}

\author{Patrick~Gill}
\affiliation{National Physical Laboratory, Hampton Road, Teddington, TW11~0LW, UK}

\begin{abstract}
We present the design, construction, and characterisation of longitudinal- and transverse-field Zeeman slowers, based on arrays of permanent magnets, for slowing thermal beams of atomic Sr.   The slowers are optimised for operation with deceleration related to the local laser intensity (by the parameter $\epsilon$), which uses more effectively the available laser power, in contrast to the usual constant deceleration mode.  Slowing efficiencies of up to $\approx$ $18$~$\%$ are realised and compared to those predicted by modelling.  We highlight the transverse-field slower, which is compact, highly tunable, light-weight, and requires no electrical power, as a simple solution to slowing Sr, well-suited to spaceborne application.  For $^{88}$Sr we achieve a slow-atom flux of around $6\times 10^9$~atoms$\,$s$^{-1}$ at $30$~ms$^{-1}$, loading approximately $5\times 10^8$ atoms in to a magneto-optical-trap (MOT), and capture all isotopes in approximate relative natural abundances.
\end{abstract}

\maketitle

\section{Introduction}

Production of ultracold atomic gases in the laboratory provides a unique opportunity for the exploration and application of many-body quantum systems.  Due to a rich two-electron structure, divalent elements such as Sr and Yb are receiving particular attention.  The realisation of ultra-precise optical clocks \cite{Katori2002, Campbell2008, Falke2011, Hong2009, Baillard2008} also drives use of laser-cooled Sr in various precision measurement experiments, e.g. to determine local gravity \cite{Poli2011}, form the basis of extremely narrow linewidth lasers \cite{Meiser2009}, provide a route towards quantum computing \cite{Daley2008}, and perform Rydberg spectroscopy in ultracold plasmas \cite{Millen2011}.  Bose-Einstein condensation (BEC) of $^{84}$Sr and $^{86}$Sr, and degenerate Fermi gases of $^{87}$Sr have been recently achieved  \cite{Stellmer2009, Escobar2009, Stellmer2010, DeSalvo2010, Tey2010}, providing a platform for a wealth of interesting studies.  Furthermore, the production of ultracold Sr$_2$ molecules \cite{Reinaudi2012, Stellmer2012} finds applications in molecular clocks, which may be used to investigate possible time variation of the electron-proton mass ratio \cite{Kotochigova2009}.   

An essential first step in the production of these ultracold atomic gases is rapid loading of a magneto-optical trap (MOT).  A high flux of slow atoms is therefore required, which is often achieved with an atomic beam apparatus exploiting the Zeeman slowing technique \cite{Phillips1982}.  Alternative magneto-optical schemes \cite{Ovchinnikov2005} loaded from thermal vapors typically provide superior mono-energetic beams but are less straightforward to implement effectively in alkaline earths.  Nevertheless, such sources are under development \cite{Bongs2012, Jones2012}.  

Conventional Zeeman slowers use current-carrying tapered solenoids to generate the required magnetic field for compensation of the varying Doppler shift experienced during slowing \cite{Courtillot2003}.  Such slowers often require a significant current, and therefore large power supply, and, in some cases, the added complication and mechanical noise of water cooling.  In addition, the field profile can be difficult to tune, and the coil windings fully enclose the atomic beam tube, restricting both optical access and the ability to remove the slower from the vacuum apparatus without disassembly.  

Recently, there have been proposals for alternative designs which circumvent these issues, using permanent magnets to generate the required field distribution \cite{Ovchinnikov2007, Ovchinnikov2012, Cheiney2011}.  The proposal of Ref.~\cite{Ovchinnikov2007} is for a transverse-field (TF) Zeeman slower consisting simply of a 2D array of permanent magnetic dipoles, positioned appropriately either side of the atomic beam tube.  Owing to the simplicity of the design, a dynamically configurable implementation of this slower was recently demonstrated for Sr \cite{Reinaudi2012a}.  TF slower designs for Rb have been previously demonstrated \cite{Melentiev2004, Cheiney2011} and benefit from a simpler realisation of the required magnetic field compared to longitudinal-field (LF) slowers.  However, this simplicity comes at the cost of doubling the necessary laser power due to the requirement of linear polarisation orthogonal to the $B$-field direction, of which only one $\sigma$-component is tuned to resonance.   

In this paper we characterise the performance of slowers for Sr based on arrays of permanent magnets with both TF and LF configurations which were developed for the optical lattice clock apparatus at the National Physical Laboratory, UK.  We begin with a general introduction to Zeeman slowing followed by a discussion of details relevant to slowing Sr and a detailed description of the construction of both the TF and LF Zeeman slowers.  The operation of the TF slower is explored in detail using measurements of slow-atom flux provided by both MOT loading and Doppler spectroscopy of longitudinal velocity distributions, and comparisons drawn with a Monte-Carlo simulation of multiple atom trajectories.  Finally, both the longitudinal and transverse efficiencies of the TF Zeeman slower are discussed and modelled. 

\section{Zeeman Slowing Strontium Atoms}

\subsection{General theory of Zeeman slowing}

The Zeeman slowing technique exploits a spatially varying magnetic field to compensate the changing Doppler shift experienced by an atom decelerated in a counterpropagating light field.  

We start by considering the velocity dependent frequency detuning, $\Delta_{\mathrm{eff}}(z)$, of an atomic resonance, $\omega_0$, and include the frequency shift due to the Zeeman effect in a magnetic field ${B}(z)$, with atomic beam propagating along $z$, counter to a laser beam at frequency $\omega_L$, such that,
\begin{equation} \label{effdet1}
\Delta_{\mathrm{eff}}(z)=\Delta_L + {{k}}{v}(z) - \mu' {B}(z)/\hbar
\end{equation}
where $\Delta_L=\omega_L-\omega_0$ is the laser detuning, ${k}$ is the wavenumber considered along $z$, $v(z)$ is the atom velocity along $z$, $\mu'=(g_eM_{e}-g_gM_g)\mu_B$ is the transition magnetic moment, $g_{g,e}$ are the Land\'{e} g-factors of the ground and excited states, $M_{g,e}$ are the magnetic quantum numbers, and $\mu_B={e\hbar}/{2m_e}$ is the Bohr magneton, where $m_e$ is the electron mass.  We define ${v}_{\mathrm{res}}(z)$ as the atom velocity given by the case of $\Delta_{\mathrm{eff}}(z)=0$.  The atom-light interaction results in a scattering force
\begin{equation}\label{scf}
{F}(v,z)=\frac{\hbar {k}\Gamma}{2}\frac{s(z)}{1+s(z)+(2\Delta_{\mathrm{eff}}(z)/\Gamma)^2},
\end{equation}
where $\Gamma$ is the photon scattering rate, and $s(z)=I(z)/I_{\mathrm{sat}}$ is the local saturation parameter, with $I(z)$ the intensity at position $z$ and $I_{\mathrm{sat}}$ the saturation intensity.  The force is maximum for $\Delta_{\mathrm{eff}}(z)=0$, and the on-resonance deceleration for an atom of mass $m_{\mathrm{a}}$ given by
\begin{equation}\label{ares}
{a}_{\mathrm{res}}(z)=\frac{\hbar {k}\Gamma}{2m_{\mathrm{a}}}\frac{s(z)}{1+s(z)}.
\end{equation}
For a slower with constant deceleration $a_{\mathrm{res}}$, this maximum deceleration imposes a lower bound to the length of Zeeman slower for a given range of slowing, as $L_{\mathrm{min}}=(v_{\mathrm{f}}^2-v_{\mathrm{i}}^2)/2a_{\mathrm{res}}$.  Stable deceleration occurs at a fraction of ${a}_{\mathrm{res}}(z)$, characterised by the $\epsilon$-parameter
\begin{equation}\label{epsilon}
\epsilon=\frac{{a}(z)}{{a}_{\mathrm{res}}(z)}\leq1,
\end{equation}
which describes the ratio of reduced local deceleration $a(z)$ to the on-resonance maximum deceleration at the local saturation $s(z)$.  
Atoms arriving at a position $z$ above $v_{\mathrm{res}}(z)$ experience a blue detuned slowing beam and a correspondingly reduced deceleration.  The extended process of slowing is thus unstable.  For $\epsilon<1$, local variations in atom velocities are compensated by corresponding variations in the deceleration, such that atoms remain on the low velocity wing of the Lorentzian scattering force profile with an equilibrium velocity ${v}_{\mathrm{eq}}(z)$.  Such an equilibrium velocity accounts for both variations in the technical implementation of the slower, such as fluctuations in the cooling light and magnetic field gradient, and also fundamental fluctuations from photon recoil events, in the direction of $z$, which are averaged over $4\pi$.  The effective operating detuning $\Delta_{\mathrm{eff}}(z)$ which is attributed to $\epsilon$ can be shown to be
\begin{equation}\label{dep}
\Delta_\epsilon(z)={k}({v}_{\mathrm{eq}}(z)-{v}_{\mathrm{res}}(z))=-\frac{\Gamma}{2}\sqrt{(1+s(z))\frac{1-\epsilon}{\epsilon}},
\end{equation}
and is equal to zero at $\epsilon=1$.  Under normal conditions, atoms arriving at a point in the slower above the equilibrium velocity experience an increased scattering force and therefore greater deceleration, and conversely slower atoms are decelerated less.  This results in a damping of the atomic motion towards ${v}_{\mathrm{eq}}(z)$ which acts to bunch or compress the velocity distribution of the slowed atoms.  This damping is maximum for an offset of ${v}_{\mathrm{eq}}(z)$ from ${v}_{\mathrm{res}}(z)$ corresponding to the maximum gradient of the scattering force, which is achieved for $\epsilon=0.75$ ($s=2$).

Finally, the velocity of atoms for a given set of parameters is determined as,
\begin{equation}\label{veq}
{v}_{\mathrm{eq}}(z)=(\Delta_{\epsilon}(z)+\Delta_B(z)-\Delta_L)/{k}
\end{equation}
where $\Delta_\epsilon$ and $\Delta_L$ are previously defined, and $\Delta_B=\mu' B_{\mathrm{T}}(z)/\hbar$ is the detuning due to the local field, $B_{\mathrm{T}}(z)=B(z)+B_{\mathrm{B}}$, with $B(z)$ the field contribution due to the slowing profile and $B_{\mathrm{B}}$ that of a uniform bias field.

In other treatments \cite{Napolitano1990, Molenaar1997}, a similar parameter $\eta$ is introduced which relates the local deceleration to the on-resonance acceleration at infinite laser intensity, i.e., for ${a}_{\mathrm{res}}(z)={\hbar {k}\Gamma}/{2m_{\mathrm{a}}}$, which is a good approximation only for slowers operated at large $s$, such as those for Rb or Na, which have moderately low saturation intensities.  The slower is then designed for a constant deceleration.  For the case of Sr, the $^1$S$_0\rightarrow\,^1$P$_1$ cooling transition has a saturation intensity of 40~mW$\,$cm$^{-2}$ and at {461}~nm it is not practical to achieve a laser power sufficient to obtain high operational values of $s$.  In this case we expect $s\simeq 1$ and therefore $\epsilon$ is a more suitable parameter.  Furthermore, this definition allows for a non-uniform deceleration throughout the slower by the included spatial dependence of the slowing beam intensity.  Efficient use of the limited slowing light is then achieved by focusing of the slowing beam to overlap optimally with the divergent atomic beam, and the effects of absorption can be included.  

\subsection{Optimum field profile}

To tune the slower for constant $\epsilon$, the optimum magnetic field profile is calculated according to equation~\eqref{veq}, where $v_{\mathrm{eq}}(z)$ is given by the solution to,
\begin{equation}
\frac{\mathrm{d}v_{\mathrm{eq}}(z)}{\mathrm{d}z}=\epsilon\frac{1}{v_{\mathrm{eq}}(z)}\frac{\hbar k \Gamma}{2 m_{\mathrm{a}}}\frac{s(z)}{1+s(z)}
\end{equation}
which is solved numerically within the intended capture and exit velocities of the slower.  For efficient use of the slowing laser power, the slowing beam is convergent towards the source of the atomic beam, providing an increased saturation, $s(z)$, and thus local deceleration, towards the slower entrance.  Counter to this, the laser power is reduced throughout the slower, towards the slower entrance, by absorption of the slowing light by the resonant atomic flux.  Here, for a typical atomic beam intensity of $\sim1.6\times 10^{14}$~atoms$\,$s$^{-1}\,$sr$^{-1}$, this reduction in power is estimated to be $\sim10$~$\%$ which is sufficiently small to be neglected in calculation of the field profile.  Further details regarding the inclusion of absorption can be found in Ref. \cite{Ovchinnikov2007}.

For a slower designed to operate at a given design $\epsilon$, $\epsilon_{\mathrm{d}}$, at a design saturation $s_{\mathrm{d}}(z)$, we may define a minimum saturation of the cooling light in 1D for which stable deceleration occurs as
\begin{equation}\label{smin}
s_{\mathrm{min}}(z)>\frac{\epsilon_{\mathrm{d}} s_{\mathrm{d}}(z)}{1+s_{\mathrm{d}}(z)(1-\epsilon_{\mathrm{d}})},
\end{equation}
which is derived from equations \eqref{ares} and \eqref{epsilon} by considering the required acceleration, ${a}(z)$, as fixed by the initial choice of $\epsilon$ and ${a}_{\mathrm{res}}(z)$.  By lowering $s(z)$, and therefore ${a}_{\mathrm{res}}(z)$, in a fixed slowing configuration, we are effectively increasing $\epsilon$ which must be bound to $<1$.  

To account for a non-uniform transverse intensity distribution of the slowing beam, which is typically Gaussian, we may generalise equation~\eqref{smin} to include the transverse beam dimension $\rho=(x^2+y^2)^{1/2}$ such that $s(z,\rho)=s(z,0)\exp[-2\rho^2/w_L(z)^2]$, where $w_L(z)$ is the slowing beam waist at position $z$ and $s(z,0)$ is the on-axis saturation parameter.  Therefore, to satisfy $s(z,\rho)>s_{\mathrm{d}}(z,\rho)$ to the FWHM of the slowing beam we operate with laser intensity $s(z,0)=2s_{\mathrm{d}}(z,\rho)$, and $s_{\mathrm{d}}(z,\rho)$ corresponds to the average intensity of the Gaussian beam.

\subsection{Zeeman shift of the $^1$S$_0$ and $^1$P$_1$ states}\label{srlevels}

\begin{figure}
\centering
\includegraphics[width=2.5in]{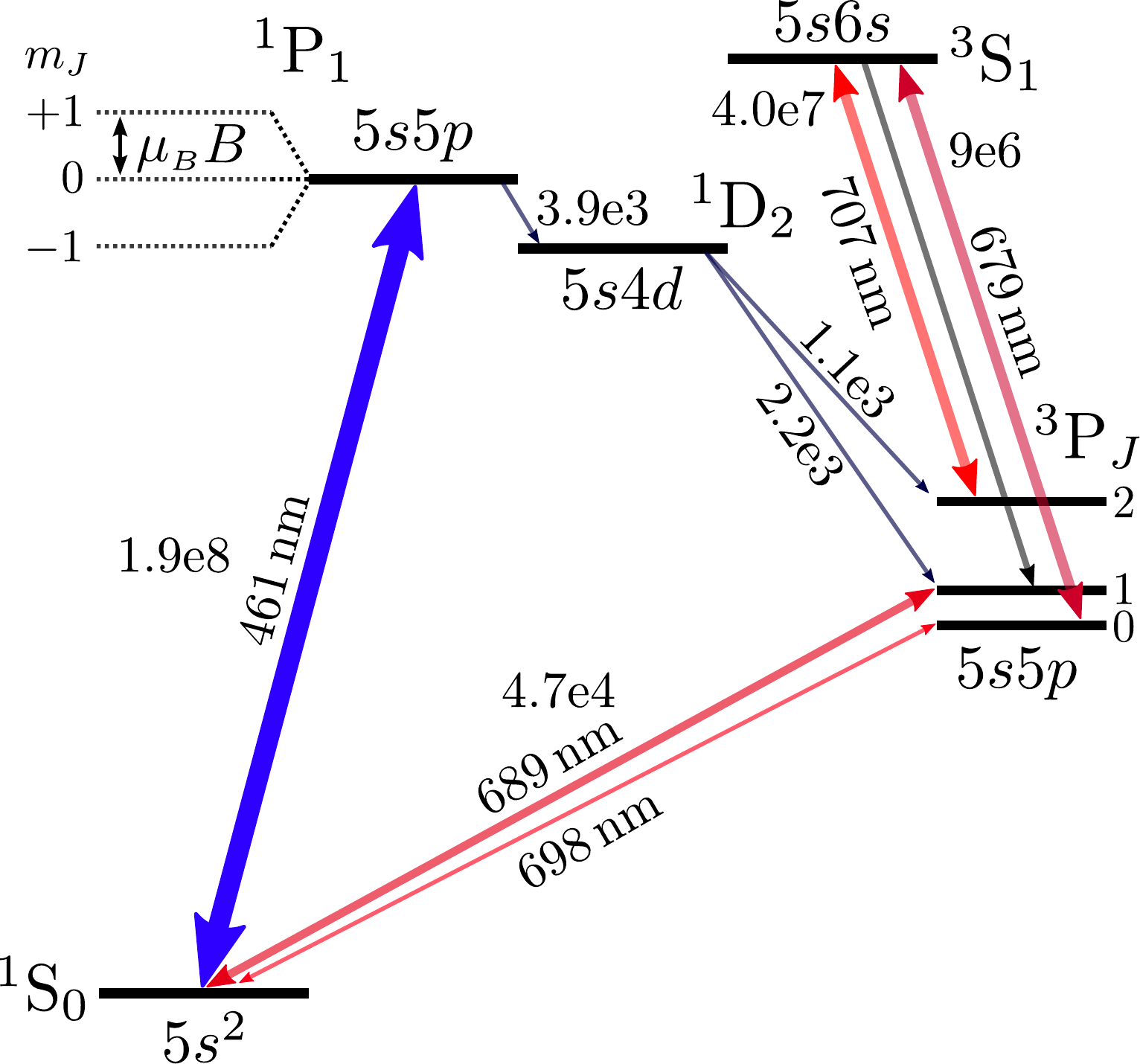}
\caption{Energy levels used in laser cooling Sr, shown for the bosonic isotopes.  Decay rates, given adjacent to each transition where appropriate, are in units of $s^{-1}$.}
\label{fig_levels}
\end{figure}

The level structure relevant to laser cooling Sr is shown in Fig$.$~\ref{fig_levels}.  The slower operates on the $^1$S$_0\leftrightarrow^1$P$_1$ transition at  461~nm, which has a photon scattering rate of $\Gamma={1.90\times10^8}$~s$^{-1}$ allowing efficient slowing over relatively short lengths of a few tens of centimetres. 

For the bosonic isotopes of Sr, with zero nuclear spin, the $^1$S$_0$ ground state has Land\'{e} g-factor $g_J=0$, and so no first-order Zeeman shift occurs.  The $^1$P$_1$ excited state has $g_J=1$ and magnetic substates $M_J=0, \pm 1$, which are shifted in energy, $\Delta E_M$, by the applied field, $B_z$, according to $\Delta E_M=\mu_B g_J M_J B_z$, where $\mu_B$ is the Bohr magneton.  The resulting frequency shift of the $^1$S$_0\leftrightarrow^1$P$_1$ transition is $\Delta\nu\simeq 1.4\,M_J$~$\mathrm{MHz/Gauss}$.


The situation for the fermionic isotope, $^{87}$Sr, with nuclear spin $I=9/2$, is a little more complex owing to the additional hyperfine interaction.  The $^1$P$_1$ excited state is split into 3 hyperfine components with total angular momentum $F=7/2, 9/2$, and $11/2$.  For slowing and cooling we operate on the $F\rightarrow F+1$ transition from the $\left|F=9/2\right>$ ground state to the stretched $\left|F=11/2, M_F=-11/2\right>$ excited state.  Here, calculation of the appropriate $g$-factors must include the nuclear $g$-factor, $g_I$, which is a small number given by $g_I={\mu_I(1-\sigma_d)}/({\mu_B|I|})$, where $\mu_I=-1.0924\mu_N$ is the nuclear magnetic moment and $\sigma_d=0.00345$ the diamagnetic correction \cite{Boyd2007},  $\mu_N={e\hbar}/{2m_p}$ is the nuclear magneton, and $m_p$ the proton mass.   For the $^1$P$_1$ state $g_F\simeq2/11$, which provides a Zeeman shift of the stretched state equivalent to the bosonic case.  A Zeeman splitting for the $^1$S$_0$ $\left|F=9/2, M_F=9/2\right>$ ground state of $\Delta\nu_g\simeq8.3$~kHz$\,$(mT)$^{-1}$ is small enough in comparison to the $^1$S$_0\leftrightarrow^1$P$_1$ transition linewidth ($\gamma=30.2$~MHz) to negate any effects of optical pumping and is ignored.

\section{Zeeman Slowers based on Permanent Magnetic Dipoles}

The slowers described here are based on arrays of permanent magnetic dipoles (MD), as proposed in Ref.~\cite{Ovchinnikov2007, Ovchinnikov2012}.  Typically we operate with $\epsilon$ between 0.6 and 0.7, capture velocity $v_{\mathrm{c}}\geq 400$~ms$^{-1}$ and exit velocity $v_{\mathrm{f}}\simeq30$~ms$^{-1}$, and slowing beam powers $P_L$ up to 80~mW focussed to overlap well with the atomic beam.  The field profile varies over a range of $\sim60$~mT, which is implemented from approximately $-30$~mT to $+30$~mT to minimise the maximum magnitude of field which must be produced.  The exit field of 30~mT also ensures the slowing beam detuning is sufficiently far from resonance with atoms extracted from the slower exit \footnote{We have also tested single polarity slowers with 0~G exit field but without MOT loading.  In this way the exit of the slower is effectively moved towards the MOT region (decreasing divergence) and post-cooling forms an integral part of the slower design, however extraction of the slow atoms is not well-defined and the resonant light will likely be problematic for operation of the MOT.}, such that the force due to the slowing beam $F_{\mathrm{SB}}$ is small compared to that of the MOT beams, $F_{\mathrm{MOT}}$.  Here, $F_{\mathrm{SB}}/F_{\mathrm{MOT}}<0.1$, and MOT operation is relatively unimpeded.


\subsection{Realising the field with magnetic dipoles}

A simple point-like dipole model, confirmed by a finite element method model with a finite size of magnet  \cite{Ovchinnikov2005}, is sufficient to calculate the field distribution near the axis of the Zeeman slower.

The magnetic flux density from a point-like magnetic dipole placed at the origin of a Cartesian coordinate system and oriented along the $x$-axis is described by the following set of equations,
\begin{align}
B_x=&\frac{\mu_0 m}{4\pi}\left[\frac{2x^2-y^2-z^2}{r^5}\right]\label{Bxd}\\
B_y=&\frac{\mu_0 m}{4\pi}\left[\frac{3xy}{r^5}\right]\label{Byd}\\
B_z=&\frac{\mu_0 m}{4\pi}\left[\frac{3xz}{r^5}\right]\label{Bzd},
\end{align}
where $r=({x^2+y^2+z^2})^{1/2}$, $\mu_0$ is the vacuum permeability ($\mu_0={4\pi\times 10^{-7}}$), and $m$ is the magnetic moment.

\subsubsection{Transverse-Field (TF) MD Zeeman Slower}

A suitable Zeeman slowing field profile, transverse to the atomic beam direction $z$, is produced by a 2D array of transversely oriented magnetic dipoles arranged at equal intervals $\delta z$ at positions $z_i$ and $y=0$ along the atomic beam tube, and with symmetrical distribution at various distances of $x_i$ (see figure~\ref{TFslower}) \cite{Ovchinnikov2005}.  The spatial distribution of the resulting magnetic field is determined by the summation of the individual dipole field amplitudes, which for the $B_x$ component is expressed as,
\begin{align}
B_x&=\frac{\mu_0 m}{4\pi}\sum^N_{i=1}\left[\frac{2(x-x_i)^2-y^2-(z-z_i)^2}{((x-x_i)^2+y^2+(z-z_i)^2)^{5/2}}\right].
\end{align}
Along the slower axis the contribution from components $B_y$ and $B_z$ each cancel due to symmetry and $B_x$ determines the slowing profile.  Away from the axis, components $B_y$ and $B_z$ are computed in a similar manner and summed to calculate the field at an arbitrary position.  Deviation from the slower axis sees an undesirable transverse field curvature along $x$ and $y$, which is greatest along $x$ and at regions of high field.  For the implementation presented here (figure~\ref{TFslower}), the variation across the slower exit provides a symmetrical $\sim1.5$~mT ($\approx5~\%$) increase at the transverse extremities of the atomic beam tube ($\pm$~8~mm from axial centre), which corresponds to a maximum variation of exit velocity of 5~ms$^{-1}$.  Such variations should be accounted for by conservative choice of $\epsilon$, and are minimised, for a given desired field, by use of a higher strength magnet positioned further from the atomic beam tube.  However, when aiming for a rapid field decay at the slower exit, a smaller magnet positioned closer to the beam tube is preferable.  Due to a non-ideal implementation of the field, a small longitudinal field contribution, $\sim10~\%$ of the transverse field, is measured along the slower axis.  Both effects are small and do not significantly affect the slowing dynamics.

\subsubsection{Longitudinal-Field (LF) MD Zeeman Slower}

A suitable longitudinal magnetic field profile, mimicking that of a tapered solenoid, may be produced by an array of magnetic dipoles oriented longitudinally along $z$ with the dipole axes $x$ and $z$ exchanged in equations \eqref{Bxd},\eqref{Byd},\eqref{Bzd}.  The axial field, which defines the slowing profile, is thus,
\begin{align}
B_z=&\frac{\mu_0 m}{4\pi}\sum^N_{i=1}\left[\frac{2(z-z_i)^2-(y-y_i)^2-(x-x_i)^2}{((x-x_i)^2+(y-y_i)^2+(z-z_i)^2)^{5/2}}\right],
\end{align}
with $B_x$ and $B_y$ components each canceling for a symmetric distribution.  Since we are using a weaker off-axis dipole field we require a 3D array with four identical columns of magnets distributed equally around a circle of radius $r_i$, with $r_i=({x_i^2+y_i^2})^{1/2}$ (see figure~\ref{LFslower}) to achieve similar fields to the TF MD Zeeman slower \cite{Ovchinnikov2012}.  

A simple implementation uses 8 columns of magnets, 4 at a radius of $r_1$ and 4 at a radius $r_2$ with opposite orientation.  Adjusting the length and radius of each set results in a suitable longitudinal field distribution running between the centres of each 4 column array.  An extension to this idea requires adjustment of $r_i$ at each magnetic position within the column, making it possible to achieve the ideal field profile over longer distances in a compact manner \cite{Ovchinnikov2012}, see figure~\ref{LFslower}.

Relative to the TF slower, the LF slower has an extended decay of the exit-field which should be compensated to improve extraction of the slow-atom flux.  For a usual solenoid slower, a compensation coil at the slower exit operating with opposite polarity is used to provide a rapid field decay.  This solution is valid here, although, with the aim of adhering to zero electrical power consumption, a suitably positioned magnetic shield is used.  

\subsection{Construction}\label{construction}

\begin{figure}[ht!]
\centering
\includegraphics[width=3.4in]{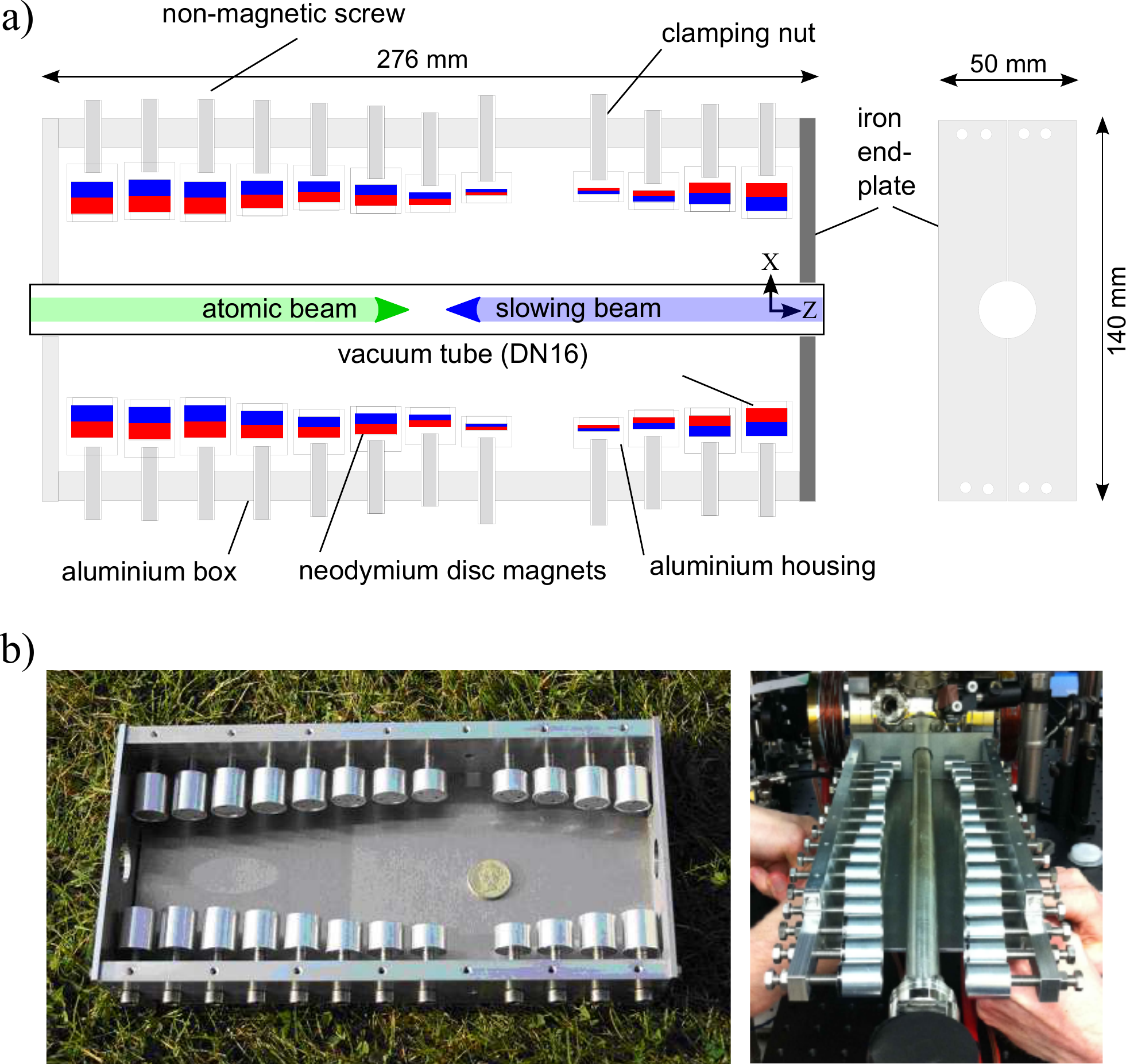}
\caption{Depiction of the transverse-field MD Zeeman slower arrangement a), and b) the constructed Sr slower developed for the EU FP7 Space Optical Clock (SOC) 2 project \cite{Schiller2012} (left), and installed extended slower (right).  Red and blue colour indicates magnet polarity.}
\label{TFslower}
\end{figure}

\begin{figure}[ht!]
\centering
\includegraphics[width=3.4in]{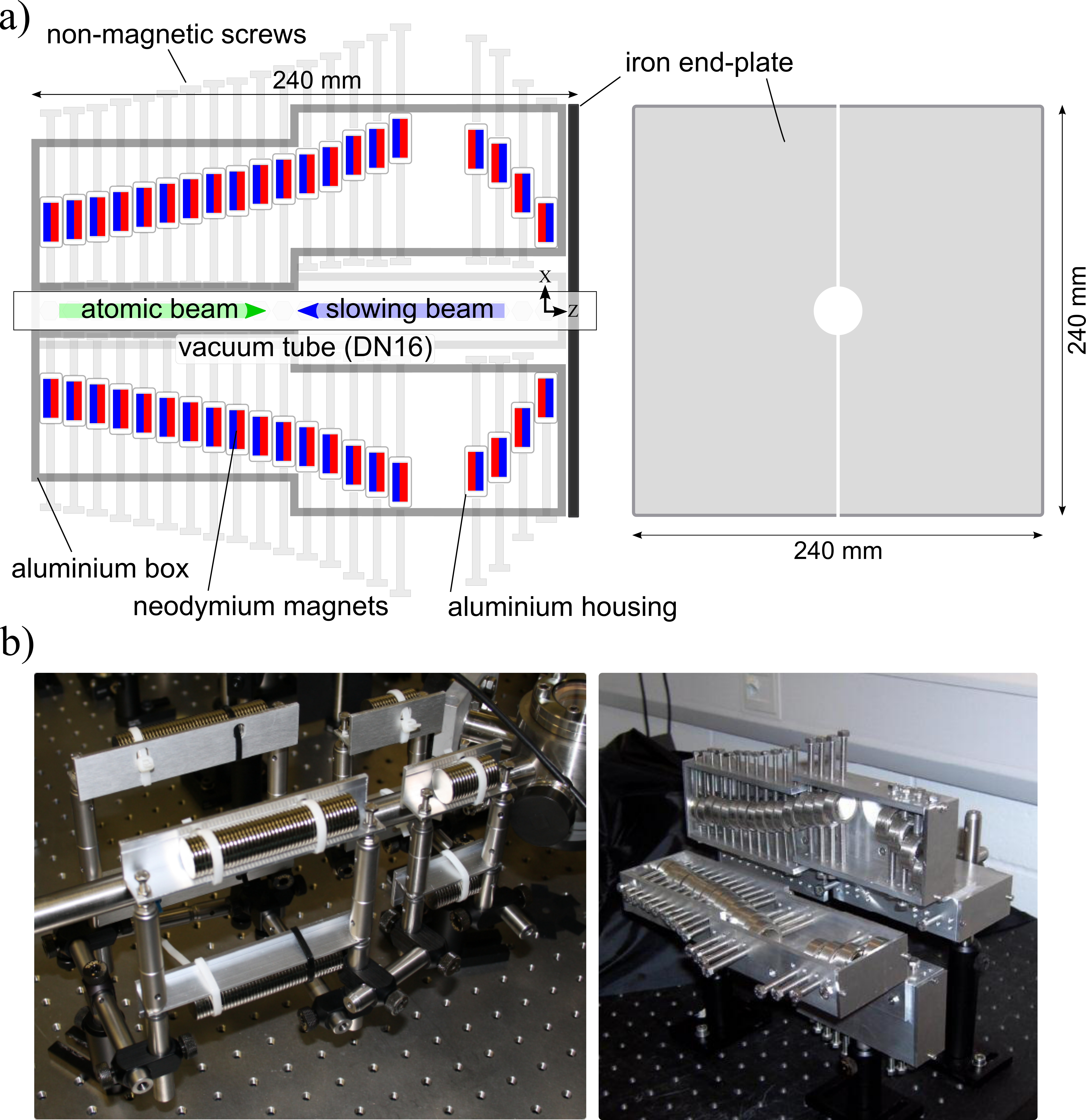}
\caption{The longitudinal-field MD Zeeman slower arrangement a), and b) pictures of LF1 (left) and LF2 (right).  Red and blue colour indicates magnet polarity.}
\label{LFslower}
\end{figure}

The dipole model of the Zeeman slower is realised using stacks of neodymium disc magnets \cite{Ovchinnikov2008}.  Neodymium iron boron (Nd$_2$Fe$_{14}$B) permanent magnets offer a high magnetic flux per unit volume, high coercivity, and are widely available at low cost.  They come in various grades such as N35, N38, N42, etc., with higher numbers generally implying a `stronger' magnet.  The strength of the magnet is determined by its residual induction (or flux density), $B_r$, and is typically of the order {1.2}~{T} for the common grade N35.  The magnetic field strength, $H$, is then given by,
\begin{equation}
H=B_r/\mu_0~~~~~~~[\mathrm{Am^{-1}}]
\end{equation}
which for a cylindrical magnet of radius $r$ and height $h$ provides a magnetic moment,
\begin{equation}
m=H\pi r^2 h~~~~~~~[\mathrm{A\cdot m^2}]
\end{equation}

The construction of the TF slower, figure~\ref{TFslower}, is extremely simple.  We use disc magnets with individual dimensions $r=7.5$~mm and $h=2$~mm, which are stacked according to the required field strength along the slower.  Each stack is housed inside a small aluminium cylindrical case which is attached to a screw thread for tuning of position $x_i$.  The magnetic dipole array is supported by an aluminium frame into which each magnet is screwed.  The only exception is at the slower exit where the frame end-plate is made from iron to provide magnetic shielding of the end-field.  Even without the shield, or any other additional compensation, the TF configuration provides a rapid decay of the magnetic field following the end-field maximum.  The addition of the shield serves to further increase the decay, and with the correct positioning also provides a slight enhancement to the end slowing field, aiding the realisation of the ideally steep field gradient at the slower exit.  The residual field contribution to the MOT region 20~cm downstream of the slower exit is $\approx16$~$\upmu$T in both $x$ and $z$ directions and $\approx2$~$\upmu$T in $y$, which is small and easily compensated.  Each TF slower is designed to operate at $\epsilon_{\mathrm{d}}=0.6$ for $s_{\mathrm{d}}(z_{\mathrm{f}})=1.5$.

We have constructed and tested two LF slowers; a simple version consisting of 8 straight columns formed of $r=7.5$~mm, $h=2$~mm stacked disc magnets (LF1, figure~\ref{LFslower} b) left), and a more involved tunable implementation using larger $r=10$~mm and $h=4$~mm magnets held 2 at a time in individual translatable housings (LF2, figure~\ref{LFslower} b) right).  For the LF2 slower, an iron shield is positioned at the slower exit to improve the extraction of slow atoms.  We make use of the additional available slowing power and design each LF slower for operation with parameters $\epsilon_{\mathrm{d}}=0.7$, and $s_{\mathrm{d}}(z_{\mathrm{f}})=4$ at the slower exit and decrease the length of the slower by $60~\%$, or 10~cm, in comparison to TF1 for a similar velocity capture range.  


For rapid tuning of each slower, we measure the magnetic field profile using a commercial gaussmeter (Bell 610 model) and transverse and axial Hall probes mounted to a motorised linear actuator, and iteratively adjust the magnet positions until the desired field profile is achieved.  The point-like dipole model provides a good starting point for the magnet positions but requires adjustment due to variations in the actual magnet strength.  The tuning process is much simpler for the TF slower due to the simple 2D arrangement, which is both mechanically trivial to adjust (by turning a screw to translate the magnet position) and simple to predict.  Tuning the LF slower requires somewhat more patience in practice.

\begin{figure}[t]
\centering
\includegraphics[width=3.4in]{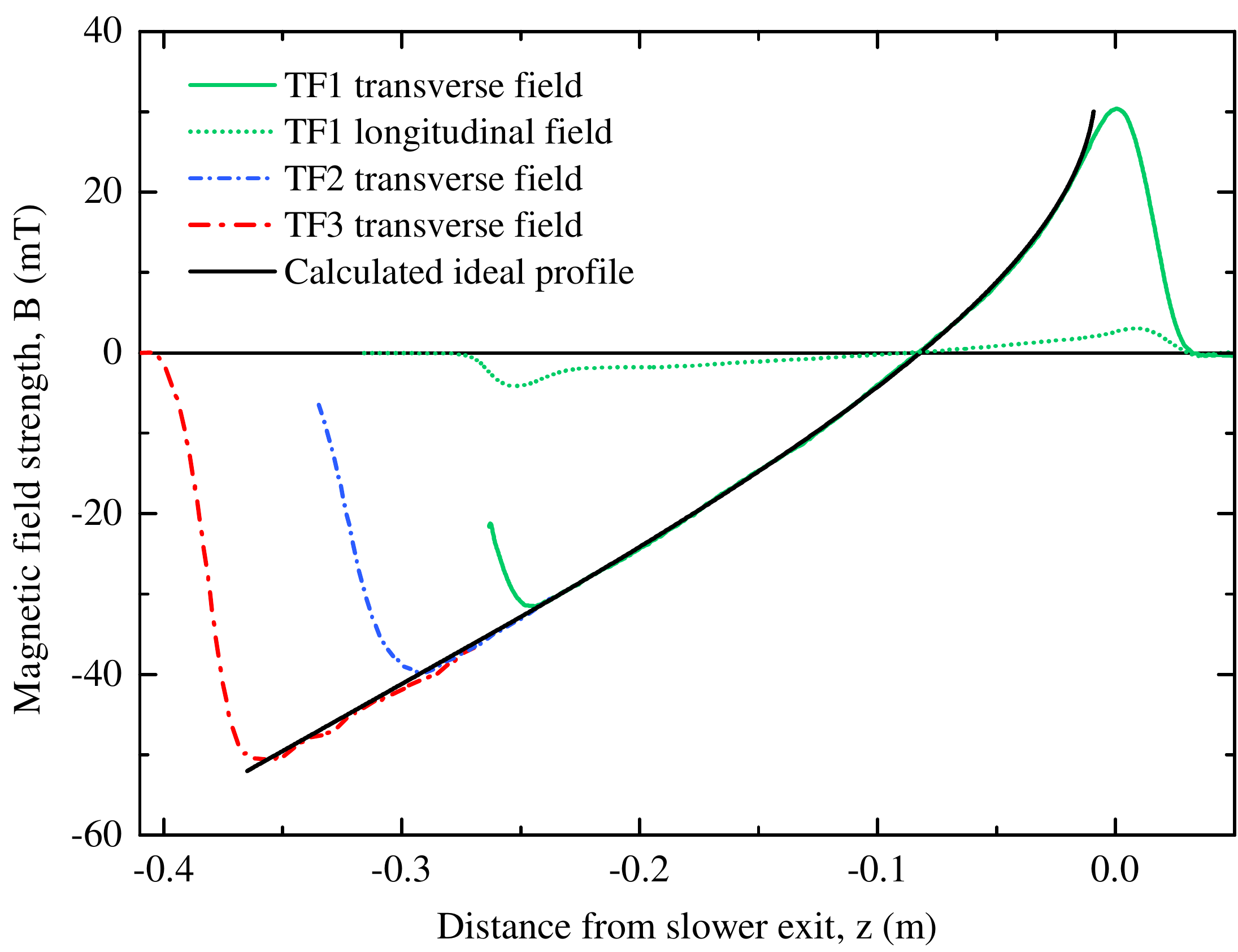}
\caption{The measured centre-line field profile of the transverse-field MD Zeeman slower in 3 lengths with the theoretical curve for a slower designed for $\epsilon_{\mathrm{d}}=0.6$ and convergent slowing beam with exit saturation $s_{\mathrm{d}}(z_{\mathrm{f}})=1.5$.  The axial longitudinal field, $<10~\%$ of the desired transverse field, is also shown (dotted-line).}
\label{fig_soc2measured}
\end{figure}

\begin{figure}[t]
\centering
\includegraphics[width=3.4in]{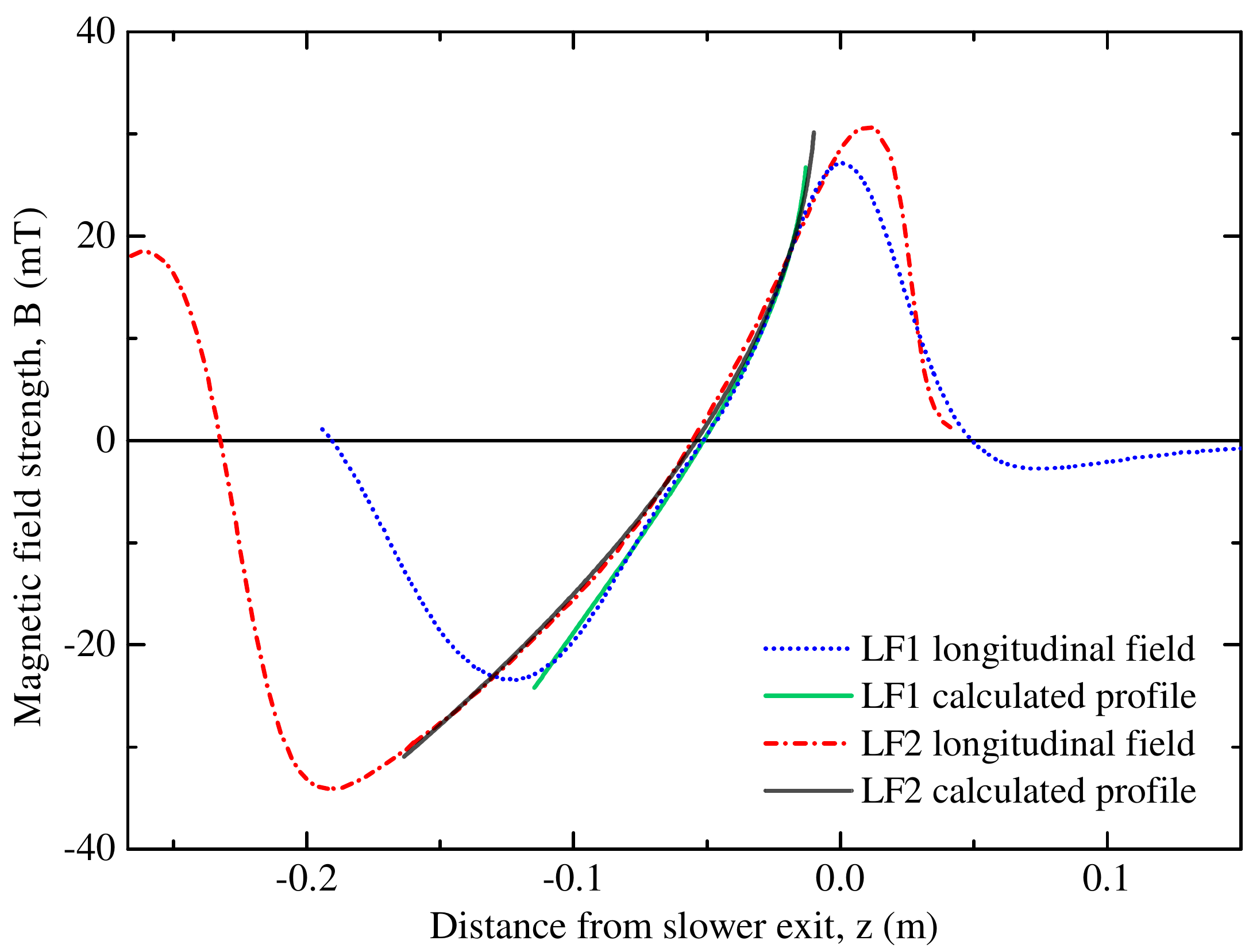}
\caption{The measured axial longitudinal field profiles for LF1 (dotted line) and LF2 (dash-dot line) longitudinal-field MD Zeeman slowers with associated intended field profiles (dashed line and solid line respectively).  Each slower is designed for $\epsilon_{\mathrm{d}}=0.7$ with a convergent slowing beam with $s_{\mathrm{d}}(z_{\mathrm{f}})=4$ at the slower exit.  The difference in field gradient towards the slower entrance accounts for a differing slowing beam geometry.}
\label{LFmeasured}
\end{figure}

Measured field profiles of the tuned TF slower are shown in figure~\ref{fig_soc2measured} for three lengths of slower; the original length (TF1) with 12 pairs of magnets, an extended version (TF2) with 3 additional magnet pairs, and a further extended version with 6 additional magnet pairs (TF3).  The axial longitudinal fields of slowers LF1 and LF2 are also measured, and shown in figure~\ref{LFmeasured}.  Further details of each slower can be found in table \ref{table_ZScomp}.  



\section{Characterisation}

The MD slowers are characterised in several ways: Doppler spectroscopy of the longitudinal velocity distributions, recorded in the MOT region downstream of the slower exit; MOT loading; and a 3D Monte Carlo semi-classical simulation.  In addition to the intended slowing, measurements here include the effects from two further interaction zones referred to as \textit{extraction}, where atoms move through the decaying field at the slower exit, and \textit{post-cooling}, which occurs in the low-field region after the slower exit.  

\subsection{Experimental setup}

The initial atomic velocity distribution is defined by the source temperature and geometry.  The atomic beam is produced by expansion through a 1~mm diameter channel of length 11~mm, which provides a high flux of $10^{14}$~atoms$\,$s$^{-1}$ in a spectroscopy chamber immediately following the Sr oven.  For further collimation, an additional 1~mm aperture is placed 10~cm downstream of the channel exit, which provides a highly collimated beam with divergence half angle $\sim10$~mrad, and typical flux $\sim10^{11}$~atoms$\,$s$^{-1}$ for a source temperature of $600\,^{\circ}$C.  The MOT region is situated 69~cm downstream.  The oven is heated by a single in-vacuum mineral-insulated resistive element wrapped around a stainless steel crucible.  Two polished stainless steel thermal shields provide good radiative isolation and allow a high efficiency of heating, requiring an electrical power of 30~W to reach a temperature of $600\,^{\circ}$C.

A commercial master oscillator power amplifier plus second-harmonic generation system provides around 350~mW at 461~nm for distribution between the MOT beams and slowing beam.  Long-term frequency stability of the 922~nm master laser is provided by stabilisation to a low-drift tunable optical reference cavity using the Pound-Drever-Hall method.  Due to a poor quality of the 461~nm spatial mode and beam pointing instability all beams are fibre-delivered to the atomic vacuum chamber through single-mode, polarisation-maintaining optical fibre and intensity stabilised.  This was particularly important for the slowing beam delivery which otherwise degraded the signal to noise in the detection region below useful levels.  We operate with a slowing beam of $1/e^2$ beam waist $w_{L}(z_{\mathrm{f}})\approx6.5$~mm and convergence half-angle 12~mrad towards the atomic source.  At the slower exit the beam size is limited by the vacuum apparatus geometry, and is intended to overlap well with the diverging atomic beam.  For TF slowers the slowing beam is polarised linearly and orthogonal to the $B$-field axis.  For LF slowers the beam has $\sigma^-$ circular polarisation.
\subsection{Measurement of longitudinal velocity distributions}


Longitudinal velocity distributions of the atomic beam are measured by CCD imaged Doppler fluorescence spectroscopy over the thermal velocity range by an additional probe laser at 461~nm tuned continuously by over 2~GHz.  The probe, with intensity well below saturation (power $\sim$~100~$\upmu$W, $1/e^2$ beam waist 1~mm, $s_0\sim0.16$), intersects the centre of the atomic beam at $55\pm1\,^{\circ}$ to the flow direction, providing a $24\pm0.7$~ms$^{-1}$ resolution and sensitivity $0.80\pm0.02$~ms$^{-1}$/MHz.  For calibration of zero velocity, a portion of the probe light is aligned through an additional cell transverse to the atomic beam propagation direction, and $^{88}$Sr resonance detected, independent of the longitudinal distribution.  A Doppler broadened feature of 34~MHz is observed, compared to the natural linewidth of 30.2~MHz, which indicates an almost natural line limited resolution of $15.7$~ms$^{-1}$.  The dominant uncertainty in peak position is derived from the angular uncertainty of the transverse probe direction, which is minimised by optimising the overlap of fluorescence peaks from the probe and its retroreflection to achieve a minimal width of feature.  Assuming a most-probable longitudinal beam velocity of 500~ms$^{-1}$, and Doppler width of similar magnitude, the observed 4~MHz broadening provides a worst case error of $\sim\pm4$~ms$^{-1}$ offset, corresponding to an angular misalignment of $<\pm0.5^\circ$.

The frequency scan of the probe light is calibrated by recording the transmission of the scanning $461$~nm probe light through a fixed length Fabry-P\'{e}rot interferometer, with free-spectral range $300$~MHz.  The calibration is checked simultaneously by spectroscopy of the known $^{88}$Sr$-^{86}$Sr isotope shift, of $-$124.5~MHz, using the transverse probe beam.  The slowing beam detuning is determined to within 2~MHz by measuring an optical beat with a resonant transverse probe beam.

\subsection{Transverse velocity effects}\label{transversevelocity}

As well as measuring the longitudinal velocity distribution, CCD imaging allows analysis of the transverse spatial distribution of the slowed atomic beam, which shows significant divergence for low longitudinal atom velocities.  This is explained by two effects: 1) {atomic beam geometry} and 2) {transverse heating}.  

First, geometrical constraints of the atomic beam aperture define an initial lower limit to the longitudinal-to-transverse velocity ratio of $v_{\mathrm{long}}/v_{\mathrm{trans}}$, $\sim10^2$ for our 10~mrad source divergence.  The slower acts predominantly to reduce $v_{\mathrm{long}}$ resulting in a decreased $v_{\mathrm{long}}/v_{\mathrm{trans}}$ and a corresponding increase in beam divergence, which is greatest for atoms experiencing a large $\Delta v_{\mathrm{long}}$.  For atoms exiting the source at $v_{\mathrm{long}}=500$~ms$^{-1}$, $v_{\mathrm{trans}}\leq$~5~ms$^{-1}$, which when slowed to $v_{\mathrm{long}}=25$~ms$^{-1}$ gives $v_{\mathrm{long}}/v_{\mathrm{trans}}=5$, or a divergence half angle of 200~mrad.  Throughout the $\sim2$~ms slowing period a transverse spread of half-width 1~cm develops, which continues in the region of free flight following the Zeeman slower exit at a rate of 2 mm$\,$cm$^{-1}$.

Second, atoms decelerated from higher initial longitudinal velocities also undergo more photon scattering events, $n_{\mathrm{sc}}(t)=(v_\mathrm{i}-v_\mathrm{f}(t))/v_r$, where $v_r$ is the recoil velocity, and are therefore subject to a greater effect of transverse heating---a result of the random recoil direction of spontaneous decay leading to a random walk in velocity space.  The root-mean-square transverse velocity $v_{x,y}$ can be estimated by $v^{\mathrm{rms}}_{\mathrm{trans}}(t)\approx v_r({ n_{\mathrm{sc}}(t)}/3)^{1/2}$ with the corresponding half-width of the transverse beam spread given by $\delta(t)=t\,v_{\mathrm{rms}}(t)$, where $t=L/v_{\mathrm{long}}^{\mathrm{rms}}$ is the flight time \cite{Joffe1993, Letokhov1981}.  Applying this to Sr, with $v_r=9.8$ mms$^{-1}$, slowing from 500~ms$^{-1}$ to 25~ms$^{-1}$, giving $n_{\mathrm{sc}}\approx50000$, results in an additional root-mean-square transverse velocity $v_{\mathrm{trans}}^{\mathrm{rms}}\approx1.2$~ms$^{-1}$ and $\delta\approx2.5$~mm at the slower exit, which continues at a rate of 0.5~mm$\,$cm$^{-1}$ in free flight.

In estimating the atom flux-density from the measured longitudinal velocity distribution, we account for the transverse spatial distribution by the scaling factor $\delta^2(v_{\mathrm{long}})$, which assumes a radially symmetric divergence.

\subsection{3D Monte Carlo simulation}\label{MC}

Our simulation is based on a first-order Euler approximation to the equations of atomic motion as governed by the local force, calculated by equation~\eqref{scf} at each time iteration for a single atom traversing the Zeeman slower.  Typically, 50000 atom trajectories are calculated to build up a representative distribution of the longitudinal atom velocities, which is extracted at a chosen region of interest.  Due to a lack of radial symmetry, we carry out a full 3D simulation using 3D point-like dipole model of the TF Zeeman slower magnetic field, which is configured to match the measured axial magnetic field distribution of the TF slower, including the regions of field outside of the intended slowing region at the slower entrance and exit.  A Gaussian slowing beam is described with geometry carefully matched to the experimental implementation.  To approximate the atomic beam source and initialise our atomic trajectory we first select the longitudinal atom velocity from a Maxwellian thermal distribution with source temperature 600~$^{\circ}$C.  This value is then used to scale a Gaussian transverse atomic velocity distribution as constrained by the oven nozzle geometry.  Atoms with larger longitudinal velocity have a correspondingly increased range of allowed transverse velocities.  We treat the centre point of the oven circular channel nozzle as the origin of all atoms.  We do not include effects of branching losses or spontaneous heating.

\subsection{Results}\label{results}

\begin{figure}[t]
\centering
\includegraphics[width=3.3in]{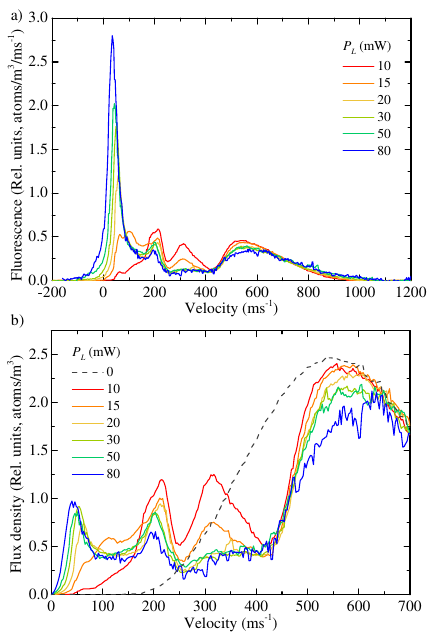}
\caption{Longitudinal velocity distributions measured for TF1 at various slowing beam powers (colour) and constant detuning $\Delta_L\approx-540$~MHz plotted as (top) fluorescence, and (bottom) atomic density.  Noise is increased for high slowing beam powers due to an increased background of scattered light from the vacuum chamber walls.  At the lowest slowing beam power we observe only the entrance of the slower and post-cooling region working effectively, resulting in a ``double-dip'' distribution, and no slow-atom peak.  Traces were taken with oven temperature $600\,^{\circ}$C and single-channel nozzle.}
\label{TFlvd}
\end{figure}

\begin{figure}[t]
\centering
\includegraphics[width=3.3in]{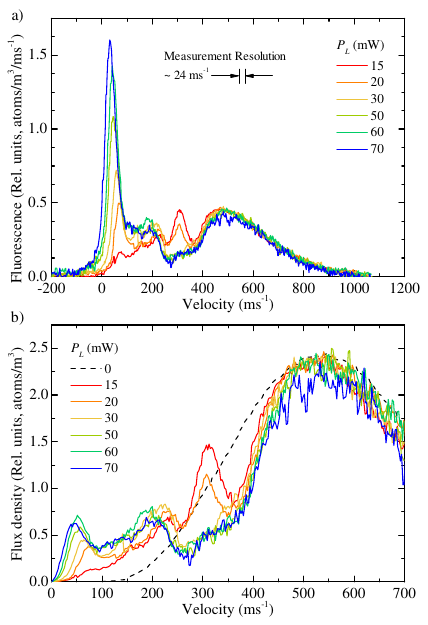}
\caption{Longitudinal velocity distributions measured for LF1 at various slowing beam powers (colour) and constant detuning $\Delta_L\approx-510$~MHz plotted as (top) fluorescence, and (bottom) atomic density.  A similar form to the TF1 velocity distribution, figure~\ref{TFlvd}, is observed for LF1 but a reduced slow-atom flux as a consequence of the smaller LF1 capture velocity.  Again, anomalous peaks are observed, at $\sim$~200~ms$^{-1}$ and $\sim$~320~ms$^{-1}$, which are pronounced at low slowing beam powers.  The post-cooling peak appears broadened in comparison to TF1 due to the nature of the exit field of LF1.}
\label{LFlvd}
\end{figure}

Longitudinal velocity distributions have been measured over a range of $P_L$ for slowers TF1 (figure~\ref{TFlvd}) and LF1 (figure~\ref{LFlvd}).  For TF1, a slow-atom flux $\approx3.5 \times 10^9$~atoms$\,$s$^{-1}$ centred at $v_{\mathrm{f}}\approx30$~ms$^{-1}$ is produced for a slowing beam detuning $\Delta_L=-540$~MHz and $P_L\approx70$~mW.  The slower is effective for laser powers down to $\approx15$~mW, as predicted by equation~\eqref{smin} for $\epsilon_{\mathrm{d}}=0.6$ and $s_{\mathrm{d}}(z_{\mathrm{f}})=1.5$, remembering only half the laser power is resonant.  At powers below this, the adiabatic following condition for atoms traversing the slower is not met, and the flux of slow atoms is substantially diminished.  To account for the Gaussian slowing beam distribution, the slower is designed to operate at an average laser intensity $s_{\mathrm{d}}(z,\rho)=s(z,0)/2=P_L/( I_{\mathrm{sat}}\pi w_L(z)^2)$.  This then corresponds to the observation of efficient operation at 30~mW.  

For the LF1 slower a similar operation to TF1 is observed.  The slower is designed to operate with saturation $s_{\mathrm{d}}(z_{\mathrm{f}})\approx4$, and utilise the full laser power which has circular polarisation.  We expect the slower to begin to operate at $\sim$~20~ mW and work efficiently at 40~mW, which is in line with observation, although it should be noted that this  is only approximate as the actual field gradient is less than the ideal towards the slower exit, requiring a lower saturation to operate.  A slow-atom flux of $2.6 \times 10^9$~atoms$\,$s$^{-1}$ is measured.  Given the reduced capture range of LF1 in comparison to TF1, we expect a reduced slow-atom flux for LF1 of approximately $40~\%$ that of TF1.  Instead, we see around $75~\%$ which points to a greater efficiency of the LF1 slower which could be due to the reduced slowing length, for which the atomic beam divergence due to the initial transverse velocity is reduced.


The expected resonant TF1 capture velocity $v_{\mathrm{c}}=440$~ms$^{-1}$ is observed as a lower limit in figure~\ref{TFlvd}.  Due to high slowing beam intensities scaling both with power and focusing of the slowing beam towards the atomic beam source, an extended interaction towards faster atoms is present, providing a somewhat extended range of $v_{\mathrm{c}}$ although to a diminishing extent.  Further, it was found that a greater slow flux is achieved for a misalignment of the slowing beam focus from the atomic beam axis, which otherwise, having $s>100$, results in a broad velocity class interaction ($\gamma'=\gamma\sqrt{s+1}\sim300$~MHz) slowing atoms both too early---they are more likely to diverge beyond the apparatus dimensions---and too far, in many cases reversing the atom velocity before Zeeman slower entry, as confirmed by Monte Carlo simulation.

The velocity distributions in figures~\ref{TFlvd} and \ref{LFlvd} show two distinct additional peaks at velocities $\sim200$~ms$^{-1}$ and $\sim310$~ms$^{-1}$, which are prominent at low slowing beam powers.  It is evident therefore that the slowing efficacy, peaking in two locations, is not absolute throughout the slowing length, and an incomplete redistribution of the thermal distribution towards the desired form results.  First, the higher velocity peak may be attributed to an extended resonant interaction at the atom entrance of the slower where the field gradient is reduced.  Subsequently, the field gradient increases and slowing can not be supported by the lower slowing beam powers.  Second, an explanation for the $\sim$~200~ms$^{-1}$ peak, termed second-peak can be given when a Doppler cooling interaction in the zero-field free-flight region following the slower exit is considered.  Provided there is some remaining distribution of thermal atoms, this post-cooling region will allow for a Doppler-tuned interaction to redistribute atoms in the second-peak, consequently burning-a-hole at higher adjacent velocities.  Such a scenario is confirmed to give similar results in our Monte Carlo simulations.  For the slower tested here, operating with $\Delta_L=-540$~MHz, the resonant capture velocity in zero field is $\approx$~245~ms$^{-1}$, as can be seen in figure~\ref{TFlvd}, and in particular for the trace corresponding to $P_L\approx10$~mW.

\begin{figure}[t!]
\raggedright
\includegraphics[width=3.4in]{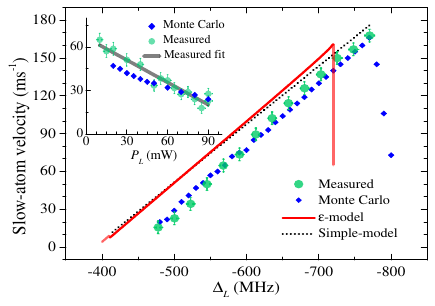}
\caption{The dependence of slow atom velocity on slowing beam detuning $\Delta_L$ and slowing beam power $P_L$ (inset) for a TF slower with exit field 26.5~mT is shown.  A final velocity of $v_{\mathrm{f}}=30$~ms$^{-1}$ for a slowing beam detuning $\Delta_L=-450$~MHz is expected but not observed due to a prolonged interaction at the slower exit.  A very good agreement is obtained between measurement and the results of a Monte-Carlo simulation which confirms an additional 30~ms$^{-1}$ of slowing occurring during extraction in a region of low field gradient at the slower exit.  The vertical line associated with the $\epsilon$-model indicates the abrupt failure of the slowing process at $\epsilon=1$.  The power dependence (inset) is taken with $\Delta_L=-520$~MHz.}
\label{vsdet}
\end{figure}

The position of the slow-atom peak has been mapped out with respect to $\Delta_L$ and is plotted together with results from different models in figure~\ref{vsdet}.  The simple-model considers only the Doppler dependence of the final velocity.  In the $\epsilon$-model this is extended to include the effective change in the $\epsilon$-parameter which results from a perceived change in the magnetic field gradient arising due to the altered working velocity range of the slower.  This adds an additional detuning dependence $\Delta_{\epsilon}$ and instructs a limit to the working velocity range of the slower, occurring for $\Delta_{\epsilon}=0$.  The models are sufficient to predict the gradient of the slow-atom velocity dependence on slowing beam detuning but do not account for the observed offset of approximately~60~MHz between the design and observed necessary $\Delta_L$, which indicates further slowing has occurred by the point of measurement.  A Monte-Carlo simulation of the atom trajectories through the slower and into a region of post-cooling reveals the origin of this effect, providing very good agreement with measurement.  During extraction of atoms from the slower, a prolonged interaction occurs around the exit-field maxima.  Due to a small field gradient in this region, a power dependent deceleration ensues and the longitudinal velocity spread increases, meaning the slow-atom peak is washed out to lower velocities on exit from the slower.  The amount of additional slowing observed gives a measure of the effectiveness of the extraction, which is desired to be abrupt.  Finally, we note that the Monte-Carlo simulation provides a cut-off velocity for stable operation of the slower in good agreement with the epsilon model, giving $v_{\mathrm{fmax}}=160$~ms$^{-1}$, but is on the boundary of our measured data set and so is not verified by experiment.

\subsection{Loading the $^1$S$_0\leftrightarrow^1$P$_1$ MOT}

For the intended slowing conditions, a slow flux $\approx3.5\times10^{9}$~atoms$\,$s$^{-1}$ centred at 30~ms$^{-1}$ is produced using TF1.  The useful slow flux exiting the slower was measured by loading of a MOT, which is operated on the transition $^1$S$_0\leftrightarrow^1$P$_1$ with parameters: detuning $\Delta=-40$~MHz, $1/e^2$ beam waist $\sim5$~mm, power $\sim3$~mW per beam (retroreflected).  We load approximately $2\times 10^{7}$ atoms into the MOT and have observed a factor of $\sim17$ enhancement with the application of repump light resonant with 707~nm and 679~nm transitions.  

In comparison to the TF slowers, LF2 was less straight forward to implement with the MOT due to the the interaction of the MOT quadrupole field with the iron end-plate of the slower, which is intended to attenuate the field beyond the slower exit to aid the extraction of slow atoms.  The effect was compensated somewhat by a permanent ring magnet positioned on the opposite side of the MOT chamber, however, a fully satisfactory solution was not completed and the measured MOT number was likely diminished as a result. We predict from measurements of the LF1 slow-atom flux and the LF2 capture velocity, a slow-atom flux of $7.1\times10^{9}$~atoms$\,$s$^{-1}$ for LF2, which is the largest of all slowers reviewed here, but this has not been confirmed experimentally.

To validate the performance of the slower for all isotopes of Sr, a continuous, synchronous scan of the slowing beam and MOT beam detunings has been carried out and fluorescence from the isotopically separated MOTs measured.  The loaded number of atoms scale approximately according to relative abundances and are separated by the relevant isotope shifts, indicating the hyperfine structure of the fermionic $^{87}$Sr atom is not problematic for the spin-flip slower configuration.

\begin{figure}[t]
\centering
\includegraphics[width=3.0in]{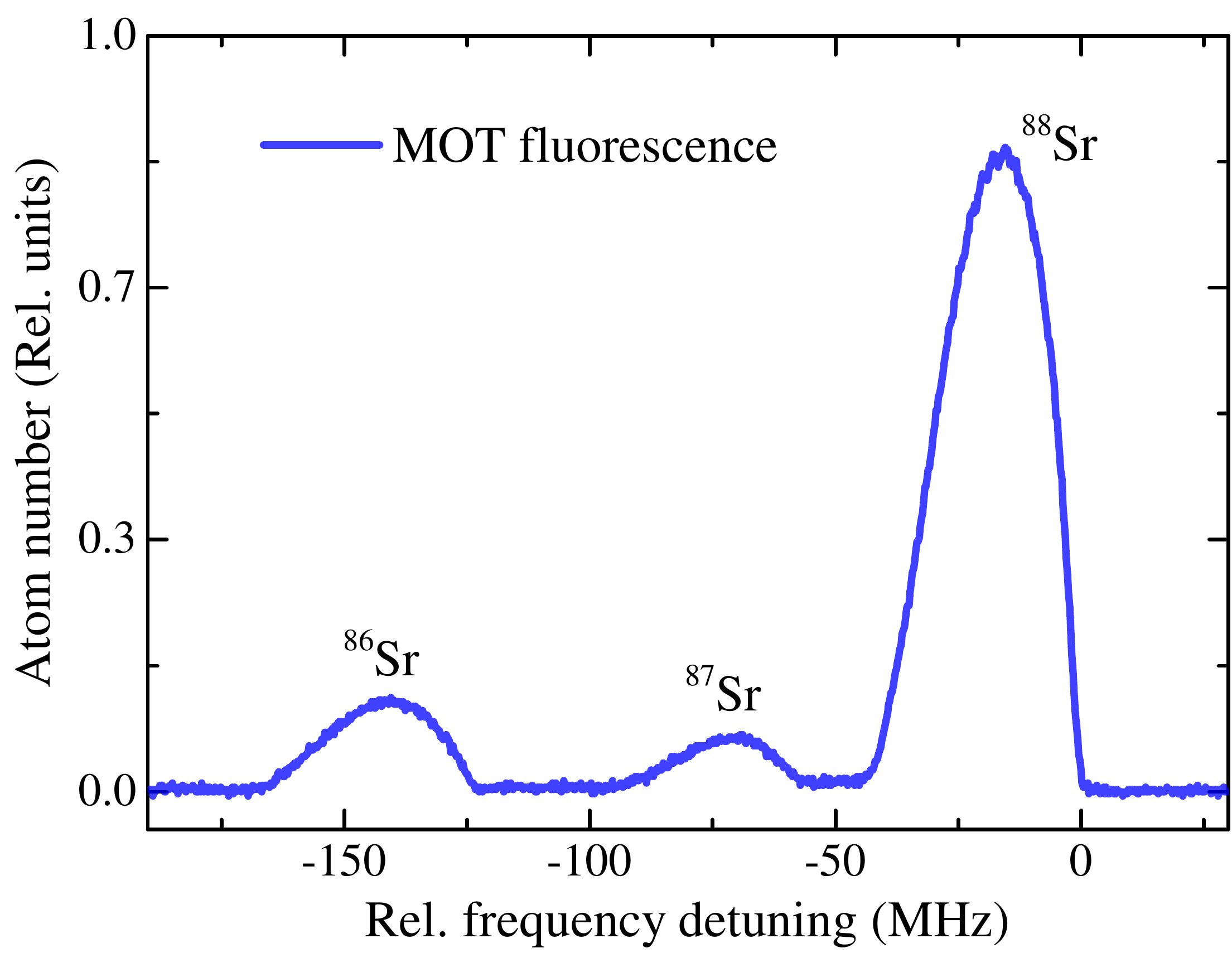}
\caption{Validation of the TF slower operation by MOT loading for isotopes (peaks left to right) $^{84}$Sr (not shown) $^{86}$Sr, $^{87}$Sr, and $^{88}$Sr.  The atom numbers are in relative agreement with natural isotopic abundance.}
\label{fig_repumps}
\end{figure}

\begin{figure}[t]
\centering
\includegraphics[width=3.2in]{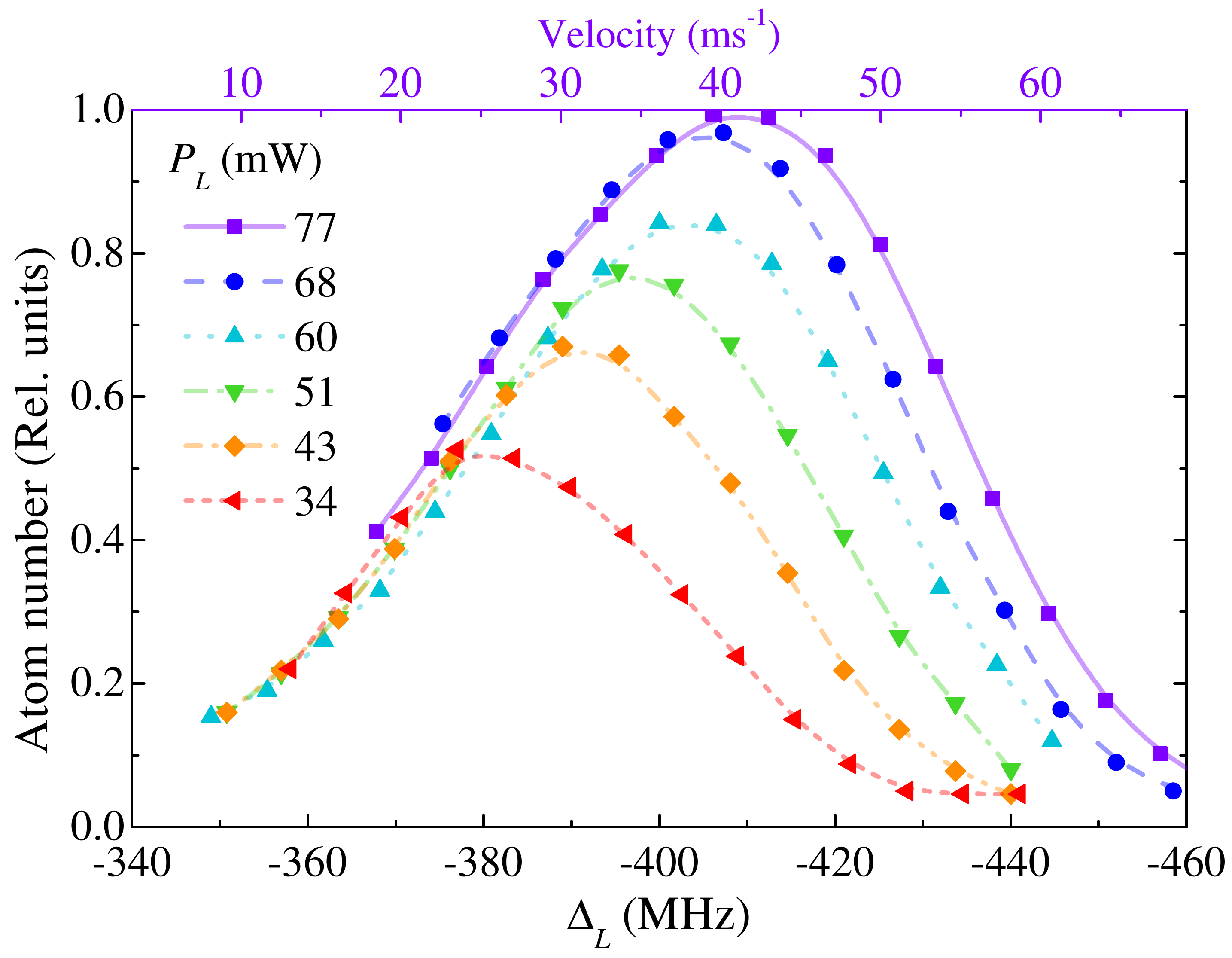}
\caption{Relative MOT atom number versus slowing beam detuning for a range of slowing beam powers.  An indication of slow atom peak velocity (top axis) is derived from a Monte Carlo simulation for the case of $P_L\approx77$~mW and not valid for the full range of powers due to the observed slow atom velocity power dependence, shown previously (figure~\ref{vsdet} b)).  The data was taken for a TF slower with exit field 18.1~mT and MOT parameters: field gradient 4~mT$\,$cm$^{-1}$, individual beam power $\sim$ 3~mW, beam diameter $\sim1$~cm, and $-40$~MHz detuning.}
\label{fig_vsbpwr}
\end{figure}

Steady-state MOT atom number (proportional to loading flux for relatively small MOT numbers $\sim10^7$ atoms), after a 0.1~s loading period without repumping lasers, has been mapped out over a range of slowing beam detunings and powers, figure~\ref{fig_vsbpwr}, from which the optimum loading conditions for a range of slowing beam powers are observed.  Owing to the power dependence of the final slow-atom velocity during extraction from the slower, a calibration of the velocity axis obtained by a measure of the slowing beam detuning dependence, described in the previous section, is not valid for the plotted range of slowing beam powers.  A trend in peak MOT loading flux towards higher $\Delta_L$, with increasing $P_L$, is observed.  To explain this, first, we acknowledge that the MOT is optimally loaded at a particular central velocity of slow atom peak.  By varying the slowing beam power we adjust both $\Delta_{\epsilon}$, see section~\ref{results}, and the amount of additional deceleration occurring during extraction of the slow atoms from the slower exit.  The measured power dependence of the slow-atom velocity (inset figure~\ref{vsdet}) is $-0.5$~ms$^{-1}$/mW and largely accounts for the approximate $-0.65(10)$~ms$^{-1}$/mW dependence of the optimum loading found here (figure~\ref{fig_vsbpwr}).  Towards lower slow-atom velocities a common decline in MOT atom number is observed.  The primary mechanism for this is an increased slow-atom divergence, see section~\ref{transversevelocity}, which reduces the density of slow flux traversing the MOT capture volume.  A 2D molasses or magneto-optical lensing phase immediately following the slower exit could be included to restore density here, and would likely improve the useful flux through the MOT capture region.

The reduction in MOT atom number at increased detunings is attributed to two processes.  First, the ultimate limit is imposed by the MOT capture velocity, $v_{c\mathrm{MOT}}$.  To gain an estimate of this we use our previous measure of the slow-atom velocity detuning dependence, figure~\ref{vsdet}, to construct a calibrated velocity scale for the data taken at $P_L\approx77$~mW, figure~\ref{fig_vsbpwr}.  From this we infer $v_{\mathrm{cMOT}}\approx50$~ms$^{-1}$, which is in agreement with a 1D model \cite{Hill2012}.  Second, it is proposed that the power dependent roll-off from this limit is due to a decreasing efficiency of the slower with increasing $\epsilon$-parameter, described in section \ref{effofslow}.

\section{Efficiency of the slower}\label{effofslow}

\begin{table*}[htbp]
\caption{Comparison of Zeeman slowers. Parentheses indicate a prediction (see text).}
\renewcommand{\arraystretch}{1.2}
\renewcommand{\tabcolsep}{0.25cm}
\label{table_ZScomp}
\centering
\begin{tabular}{lllllllll}
\toprule
Slower & $B$-field dir. & $L_0$ [m] & $s_{\mathrm{d}}(z_f)$ & $B(z_{\mathrm{c}})$ [mT] & $B(z_{\mathrm{f}})$ [mT] & $v_{\mathrm{c}}$ [ms$^{-1}$] & Slow flux [s$^{-1}$] & MOT number \\
\hline
TF1 & transverse & 0.25 & 1.5 & -31.5 & +30.4 & 463 & $3.5\times 10^9$ & $4.3\times 10^8$  \\
TF2 & transverse & 0.31 & 1.5 & -39.7 & +30.4 & 516 & $4.7\times 10^9$ & $5.8\times 10^8$  \\
TF3 & transverse & 0.38 & 1.5 & -50.6 & +30.4 & 586 & $6.1\times 10^9$ & $7.5\times 10^8$  \\
LF1 & longitudinal & 0.10 & 4.0 & -23.5 & +27.2 & 383 & $2.6\times 10^9$ & -  \\
LF2 & longitudinal & 0.15 & 4.0 & -34.1 & +30.7 & 466 & ($7.1\times 10^9$) & $3.1\times 10^8$  \\
\hline
\end{tabular}
\end{table*}



By the efficiency of a Zeeman slower we mean here the value of the fraction of the initial flux of the collimated beam of thermal atoms that are decelerated by the slower down to the designed final velocity $v_{\mathrm{f}}$.  There are several factors which influence the efficiency of a Zeeman slower for Sr atoms.


First, we make the approximation that a Zeeman slower interacts only with atoms having longitudinal velocity at or below the capture velocity $v_{\mathrm{c}}$.  Atoms with velocity greater than $v_{\mathrm{c}}$ will experience some deceleration due to the naturally broad interaction with the cooling light, however, this is typically insufficient to bring them into line with the adiabatic following condition of the field which is necessary to complete the slowing process to the slower exit.  Therefore, all atoms of the initial thermal atomic beam with velocity above $v_{\mathrm{c}}$ are discounted from contributing to the slow-atom flux exiting the Zeeman slower \footnote{There are, however, factors in the realisation of the Zeeman slower that muddy this picture.  An extended region of low field gradient at the slower entrance will serve to further this interaction to an extent that some additional portion of the thermal distribution may be slowed sufficiently to join the adiabatic following of the field further along the slower.  This effect is not considered in our presented model.}.



Second, in the process of slowing, Sr atoms are optically excited into the $5s5p\,^1$P$_1$ state which can decay to the $5s4d\,^1$D$_2$ state ($3.9\times10^3~s^{-1}$).  From here 2/3 of atoms decay to the ground state via the $5s5p\,^3$P$_1$ state ($4.7\times10^4~s^{-1}$), and the remaining 1/3 are shelved in the metastable $5s5p\,^3$P$_2$ state (see Fig.~\ref{fig_levels}).  In each case, the time for such a detour is significant compared to the total slowing time ($<2$~ms), and atoms are unlikely to return to the slowing process following this decay.  We consider these atoms as shelved in a dark state and lost from the slowing process.

Third, due to divergence of the atomic beam, its transverse size at the output of the Zeeman slower is comparable to the transverse diameter of the cooling laser beam. This leads to an additional loss of atoms from the slowing process. 

\begin{figure}[ht!]
\raggedright
\includegraphics[width=3.4in]{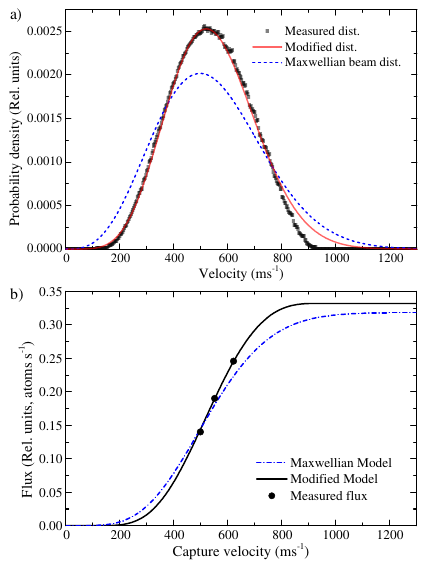}
\caption{a) Probability density distributions of the longitudinal velocity for an effusive Sr thermal (source temperature $600\,^{\circ}$C) atomic beam (dashed-line) and our measured distribution (squares), with modified Maxwellian distribution fitted (solid-line).  b) The corresponding dependency of Zeeman slowed thermal atoms on Zeeman slower capture velocity, $v_{\mathrm{c}}$, for an effusive source (dash-dot-line), and for our measured atomic beam distribution (solid-line).  The theoretical dependence is normalised with respect to total atomic beam flux, and the measured data scaled to fit.   Branching losses (BL) due to the non-closed nature of the $^1$S$_0\leftrightarrow\,^1$P$_1$ cycling transition are responsible for the saturation of the Zeeman slower slowing capacity, which here occurs at around $33~\%$ of the full atomic beam flux.  The relative slow atom flux (dots) for three lengths of TF Zeeman slower (TF1, TF2, and TF3) is inferred from steady-state MOT number.  A 50~$\%$ increase in flux is gained from TF1 to TF3 with an increase in capture velocity of around $25\%$ ($\Delta v_c\approx123$~ms$^{-1}$).}
\label{fig_fluxvc}
\label{fig_veldistr}
\end{figure}

Let us consider first losses related to the finite length of the Zeeman slower and the presence of branching losses to the $^1$D$_2$ state.  The probability of the Sr atom to stay in the ground state after scattering $n_{\mathrm{sc}}$ photons is equal to $(1-r)^{n_{\mathrm{sc}}}$, where $r=\Gamma_D/(\Gamma_S+\Gamma_D)=2.05\times10^{-5}$ is the branching ratio of a spontaneous decay of the excited state $^1$P$_1$ into the $^1$D$_2$ state ($\Gamma_D=3.9\times10^3~s^{-1}$) and the ground state $^1$S$_0$ ($\Gamma_S=1.9\times10^8~s^{-1}$).  For a Sr atom decelerated from a velocity $v$ down to velocity $v_{\mathrm{f}}$ in the slower, the number of spontaneously scattered photons is equal to $n_{\mathrm{sc}}=(v-v_{\mathrm{f}})/v_r$, where $v_r=\hbar k/m$ is one-photon recoil velocity, $k$ is the wavenumber of the scattered photon and $m$ is mass of the atom.  Taking this into account, the integrated flux of Sr atoms slowed in the Zeeman slower, normalised to the total flux of the thermal atomic beam, can be calculated as
\begin{equation}
G_L(v_{\mathrm{c}})=\frac{\int_{v_{\mathrm{f}}}^{v_c} f(v) v (1-r)^{(v-v_{\mathrm{f}})/v_r}\, \mathrm{d}v}{\int_{0}^{\infty} f(v) v\, \mathrm{d}v},
\end{equation}
where $f(v)$ is the probability density function for the longitudinal velocity distribution in the thermal atomic beam.
For an effusive thermal atomic beam (Knudsen number, $K_n=\Lambda/L\ll1$, where $L$ is the relevant length scale and $\Lambda$ is the collisional mean free path) the probability density function is given by the Maxwellian distribution $f(v)=(4/\sqrt{\pi})(v^2/u^2)\exp{(-v^2/u^2)}$, where $u=\sqrt{2k_BT/m}$.  In our experiment this is not the case.  At a source temperature of $\sim600\,^{\circ}$C the $^{88}$Sr-$^{88}$Sr collisional mean free path, $\Lambda_{88-88}$, is on the order of 100~$\mu$m.  For our circular channel nozzle of radius $r=0.5$~mm and length $l=11$~mm we have $K_n<1$ in both characteristic dimensions, $r$ and $l$, and a continuum flow regime is reached.  The corresponding measured probability density function of our Sr atomic beam is shown in Fig.~\ref{fig_veldistr}.  The fit function to the experimentally measured velocity distribution is $f(v)=C (v^4/a^5)\exp{(-v^2/a^2)}$, where $C$ is a normalisation constant and $a=328$~ms$^{-1}$ is a fit parameter.  Here we observe losses at high and low velocities resulting in a redistribution towards the mean translational energy, as is consistent with previously reported results \cite{Angel1972}.  It is noted that a multi-channel nozzle, consisting of around fifty 200~$\mu$m diameter tubes of length 10~mm, was used prior to this work.  For such a nozzle, operated with a $600\,^{\circ}$C source, the condition $l\gg\Lambda\sim r$ is satisfied, and a transition from opaque free-molecular-flow to continuum flow is observed \cite{Giordmaine1960}.  The measured velocity distribution better approximates a Maxwellian form, showing only a deficiency in low velocity atoms, however this results in an increase in the mean translational energy.

The theoretical longitudinal efficiencies of a slower, given by $G_L(v_{\mathrm{c}})$, which is a function of the capture velocity of the slower $v_{\mathrm{c}}$, for the effusive thermal atomic beam (dashed curve) and for our real velocity distribution (solid curve) are shown in figure~\ref{fig_fluxvc} b).  To verify these theoretical predictions, the flux of cold atoms produced by three lengths of TF Zeeman slower has been measured by observing loading of the MOT.  For brevity, we omit here an analysis of the LF Zeeman slower efficiency.  The measured field profiles of the tuned slowers are shown in Fig.~\ref{fig_soc2measured} and described in section~\ref{construction}.  The exit field is measured at 30.4~mT, requiring a detuning $\Delta=-570$~MHz for an exit velocity $\approx$~40~ms$^{-1}$.  The corresponding resonant capture velocities for TF1, TF2, and TF3 are given in table~\ref{table_ZScomp}.  The expected fractional increase in flux is confirmed by experiment with a gain of 35~$\%$ from TF1 to TF2, and a further 29~$\%$ increase from TF2 to TF3.  By extending TF1 by 6 magnet pairs, from 12 to 18, we achieve a factor $\sim$ 1.8 increase in slow-atom flux.  A further increase in capture velocity could potentially provide an additional factor of $\sim$~1.5 in slow-atom flux, although with diminishing returns beyond the distribution peak of $\approx$~450~ms$^{-1}$.  

The divergence of the thermal Sr atomic beam leads to an additional loss of slowed atoms. Let us consider a situation where the transverse size of the atomic beam is comparable to the transverse size of the cooling laser beam.  In the TF Zeeman slower, which is designed for a certain $\epsilon$-parameter, the atoms can be slowed only if the intensity of the laser field is higher than the critical value $I_c=2 s_{\mathrm{min}} I_{\mathrm{sat}}$, where $s_{\mathrm{min}}$ is the minimum saturation for which stable deceleration occurs given by equation \eqref{smin}, and $I_{\mathrm{sat}}$ the saturation intensity.  The factor of 2 in $I_c$ corresponds to the TF slower using only one-half of the whole laser intensity. For the LF Zeeman slower the corresponding critical intensity is $I_c=s_{\mathrm{min}}I_{\mathrm{sat}}$.  Therefore, for a Gaussian cooling laser beam with transverse intensity distribution $I(\rho)=I_0 \exp{[-2 \rho^2/w_L^2]}$, where $\rho$ is the distance from the axis of the beam, $w_L$ is the $1/e^2$ beam waist of the laser, and $I_0$ the on axis peak intensity, only the atoms with $\rho<\rho_c=w_L({\ln[2 P_L/\pi w_L^2 I_c]/2)^{1/2}}$ are fully slowed by the slower down to the final velocity.  For a Gaussian transverse distribution of the density in the atomic beam $n(\rho)=n_0 \exp{[-2 \rho^2/w_a^2]}$, the relative number of slowed atoms can be calculated as
\begin{equation}\label{KT}
G_T(P_L)=\int_{0}^{\rho_c(P_L)} \frac{4}{w_L^2} e^{-2\rho^2/w_a^2} \rho\, \mathrm{d}\rho.
\end{equation} 
The parameter $G_T(P_L)$, which is a function of the laser power $P_L$, represents the transverse efficiency of the slower.

\begin{figure}[t!]
\centering
\includegraphics[width=3.2in]{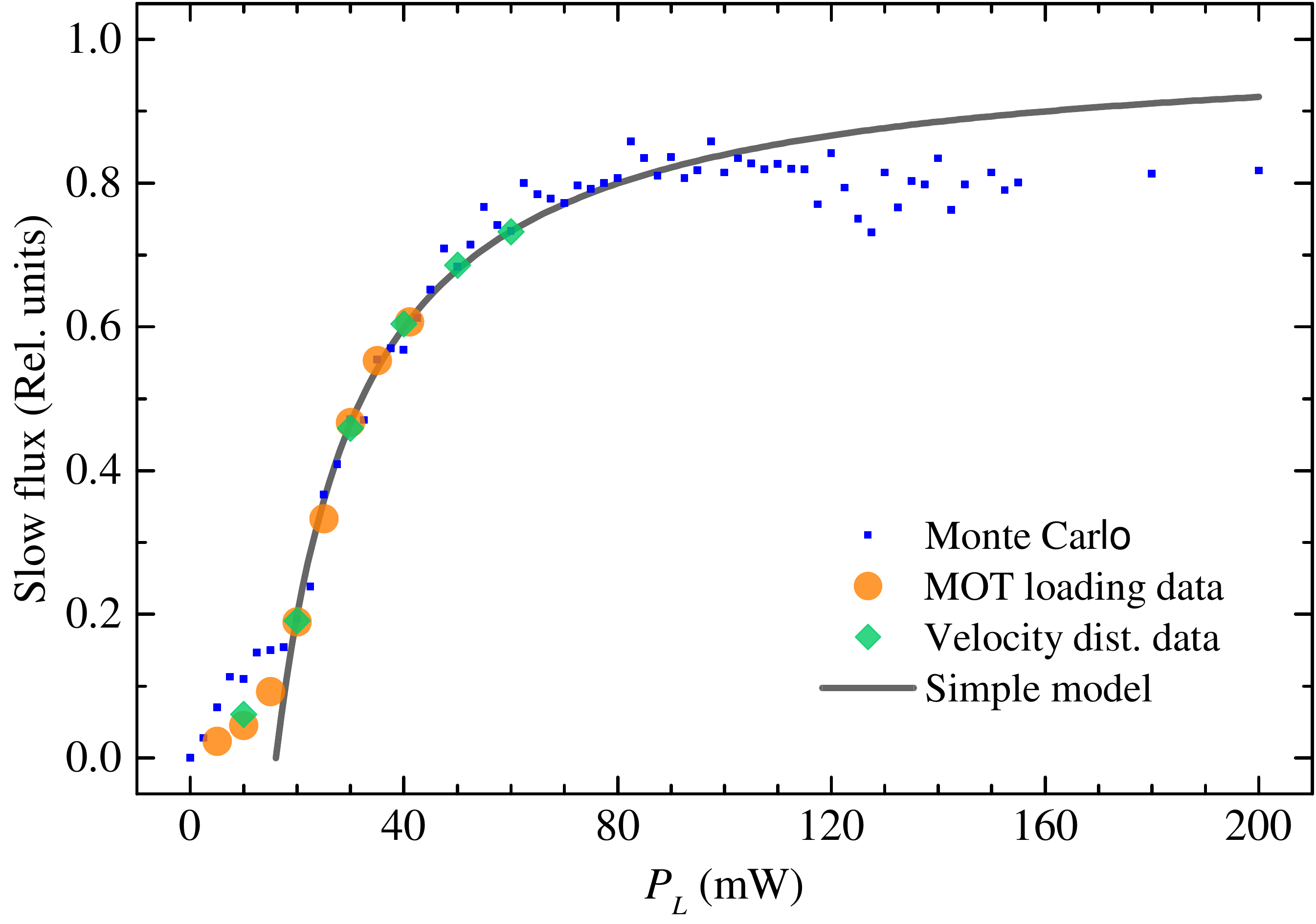}
\caption{Transverse efficiency $G(P_L)$ (solid line) of the TF1 Zeeman slower ($\epsilon_{\mathrm{d}}=0.6$, $s_{\mathrm{d}}(z_{\mathrm{f}})=1.5$) is shown in terms of relative slow-atom flux versus slowing beam power $P_L$.  Measured data are taken from longitudinal velocity distributions integrated over the slow-atom peak, and from steady-state MOT atom numbers at various $P_L$.  A $1/e^2$ slowing and atomic beam waist of $w_L=w_a=0.65$~cm at the slower exit with convergence half-angle 14~mrad is assumed throughout.  Data generated by Monte-Carlo simulation (squares) is in good agreement with the transverse efficiency model above $I_c$ and deviates only at high intensities where an extended interaction around the Zeeman slower entrance leads to a reversal of part of the atomic beam.  Below $I_c$ some flux is still present due to the region of reduced field gradient realised at the Zeeman slower exit.  This is clear for both measured data and the Monte-Carlo simulation.}
\label{fig_fluxpower}
\end{figure}

Figure~\ref{fig_fluxpower} shows experimentally measured and modeled slow-atom flux as a function of $P_L$.  Measured data is derived from the number of atoms loaded into the MOT from the TF2 Zeeman slower, and as the integrated flux from longitudinal velocity distributions. The fit line in this figure is calculated from the equation \eqref{KT} for $w_L(z_{\mathrm{f}})=w_a(z_{\mathrm{f}})=0.65$~cm and $I_c=10.18/w_L^2$~mW$\,$cm$^{-2}$. Note that, due to convergence of the cooling laser beam, its intensity and transverse radius is changing along the length of the slower.  On the other hand, for a laser beam focused to the output aperture of the atomic source, the ratio between the radii of the atomic and laser beams $w_a/w_L$ remains constant throughout the Zeeman slower.  Therefore the efficiency coefficient $G_T(P_L)$ stays the same along the length of the slower providing the slower is designed for constant $\epsilon$ parameter.  A Monte-Carlo simulation shows a good fit to the simple model above the threshold intensity (figure~\ref{fig_fluxpower}) and deviates only at high intensities where saturation of the slow flux occurs.  This behavior is attributed to the increased saturation intensity of the slowing beam at the Zeeman slower entry, which, as previously described, decelerates atoms too far in the region of the initial capture such that they do not traverse the slower.  According to the theory, there are no atoms slowed by the slower if the power of the slower is below a certain critical value (in our case $P_c\approx{17}$~mW). The small atomic flux observed below the critical power (figure~\ref{fig_fluxpower}) can be attributed to the deviation of the slowing field gradient from the ideal at the slower exit, which reduces locally $I_c$ such that slowing of thermal atoms may occur.  Such an effect is also indicated in the Monte-Carlo simulation.

Therefore, the total slowing efficiencies of Zeeman slowers TF1, TF2, and TF3, given by $G=G_T(P_L)\times G_L(v_{\mathrm{c}})$, operated with cooling laser power $P_L=70$~mW, are approximately 0.10, 0.14, and 0.18, respectively. 

Our model shows that the atomic flux of the cold atoms can be increased further by a factor approaching $1.5$ ($G=0.28$) for a slower with capture velocity $v_{\mathrm{c}}\approx800$~ms$^{-1}$ ($L_0=0.59$~m) and power of the cooling laser beam of $P_L=70$~mW for our given beam parameters.  In addition, we may estimate the fraction of slow-atom flux by integration of the flux density.  Using data from figure~\ref{TFlvd} for TF1 we estimate the slow-atom peak contains approximately 30~$\%$ of the flux captured by the Zeeman slower, implying an efficiency $G\approx0.042$.  A measured MOT loading flux $\approx3.5\times10^{9}$~$\mathrm{atoms}\,\mathrm{s^{-1}}$, from the initial atomic beam flux of $\sim1\times10^{11}$~$\mathrm{atoms}\,\mathrm{s^{-1}}$, provides a similar result.  The efficiency derived here is about a factor of two smaller than that indicated by our presented model, which can be attributed in part to an incomplete overlap of the slowing and atomic beams observed in the transverse spatial distribution.  A similar treatment may be followed for the LF Zeeman slower but has not been completed here.

\section{Outlook}\label{outlook}

In adapting the design of both TF and LF slowers to slowing other species some considerations must be made.  We take, for example, Yb which has similar scattering properties to Sr, $\gamma\approx30$~MHz and $\lambda\approx399$~nm, and approximately twice the mass.  The required slowing distance is then approximately twice that of Sr for a given value of the saturation parameter.  However, given that the most-probable velocity of a Maxwellian thermal distribution scales as ${(T/m)^{1/2}}$, where $T$ is the source temperature, and that Sr and Yb have similar vapor pressures, the required capture range and therefore length of the Yb slower is correspondingly reduced.  




For atoms with non-zero nuclear spin a Zeeman splitting of the ground state results.    For spin-flip Zeeman slowers, such as those presented in this paper, the quantisation axis is not well defined around the field zero crossing.  For large Zeeman splittings, on the order of the transition linewidth $\gamma$, or greater, a rotation of the quantisation axis within the slower can lead to optical pumping of atoms into a dark ground state where they are lost from the slowing process.  The effects of optical pumping have been observed for a permanent-magnet transverse-field spin-flip slower operated with $^7$Li atoms \cite{McClelland}.  In this case the spin-flip can simply be avoided by distributing the field in a single polarity.  


In addition to optical pumping, atoms may experience a reduced scattering force if sufficient field exists in a rogue direction to lift degeneracy of the magnetic substates.  This situation is valid in the case of a longitudinal-field slower where a single circular polarisation is chosen.  To ensure atoms are not lost from the slowing process here it may be beneficial to reduce the field gradient through the field zero to allow additional time for atoms to be slowed.  This could be implemented by producing the spin flip slower in two separated halves, each with opposite polarity.  In the transverse field configuration we require linearly polarised light, aligned orthogonal to the magnetic field direction, which is decomposed into $\sigma^{\pm}$ light.  The sensitivity of the scattering force to quantisation axis at this zero crossing is reduced by the presence of both $\sigma^+$ and $\sigma^-$ light.  For the majority of the slowing only $\sigma^-$ light is resonant with the Zeeman tuned atomic resonance, however,  interaction with the $\sigma^+$ occurs through the field zero.

The suitability of LF versus TF slower will depend somewhat on the length of slower required.  The LF slower is difficult to extend to longer lengths and is a challenge to tune.   It also requires end field compensation in the form of a coil or a shield which we find to disturb the quadrupole field of the MOT.  In its favour, the LF slower uses all the available laser power.  The TF slower configuration is well suited to longer implementations and benefits from its ease of tuning.  Effective extraction of the slow-atom flux is simple to implement and does not interfere with the MOT.  Therefore we recommend the TF slower for Sr and application to other species.

\section{Conclusion}

The design, construction, and characterisation of transverse-field (TF) and longitudinal-field (LF) permanent magnet Zeeman slowers have been described.  In particular, the TF slower apparatus is light ($\sim$~2~kg), compact, easily shielded, and consumes zero power,  which are attributes well suited to space-borne operation.  The slower also benefits from ease of tuning, which can be carried out in situ, or with the slower separated from the apparatus.  The magnet configuration allows optical access along the full length of the slower, which may be of benefit for implementing transverse cooling.  The performance of each slower is validated by measurements of the longitudinal velocity distribution, by loading of a magneto-optical-trap, and by a 3D Monte Carlo simulation.  Models of the longitudinal and transverse efficiencies of the slower are also provided.  Finally, we achieve a slow flux of $6.1\times 10^{9}$~atoms$\,$s$^{-1}$ for TF3 and an atomic beam flux of $\sim1\times10^{11}$~atoms$\,$s$^{-1}$, which compares well with other reported slowers for Sr based on conventional current-carrying solenoid designs \cite{Courtillot2003}.

\begin{acknowledgments}
The authors would like to thank Alastair Sinclair and Guido Wilpers for providing an additional laser at 461~nm for measurements of the longitudinal velocity distributions, and Rachel Godun, Richard Hobson, and Ross Williams for comments on the manuscript.  This work was funded in part by the iMERA OCS project, the EU FP7 SOC2 project, and the UK NMO programme.
\end{acknowledgments}


\bibliography{ZSreferences}

\newpage
\end{document}